\documentstyle[12pt,cite,epsfig,psfrag]{article}

\def\be{\begin{equation}}
\def\ee{\end{equation}}
\def\bea{\begin{eqnarray}}
\def\eea{\end{eqnarray}}

\newcommand{\beqal}{\begin{eqnarray}\label}
\newcommand{\beqa}{\begin{eqnarray}}
\newcommand{\eeqa}{\end{eqnarray}}

\newcommand {\f}{\frac}

\begin{document}
\baselineskip=.6cm
\begin{titlepage}
\begin{center}
\vskip .2in

{\Large \bf Brane Solutions in Time Dependent Backgrounds in D\,=\,11 Supergravity and in Type II String Theories}
\vskip .5in

{\bf Srikumar Sen Gupta \footnote{e-mail: srikumar@iopb.res.in}
}\\
\vskip .1in
{\em Institute of Physics,\\
Bhubaneswar 751 005, INDIA}

\end{center}

\begin{center} {\bf ABSTRACT}
\end{center}
\begin{quotation}\noindent
\baselineskip 15pt
We obtain explicit time dependent brane solutions in M-theory as well as in string theory by solving the reduced equations of motion (which follow, as in \cite{ASS}, from 11-d supergravity) for a class of brane solutions in curved backgrounds. The behaviour of our solutions in both asymptotic and near-horizon limits are studied. It is shown that our time dependent solutions serve as explicit examples of branes in singular, cosmological backgrounds. In some special cases the asymptotic and the boundary AdS solutions can be identified as $Milne\times R^n$ spacetime.  
\end{quotation}
\vskip 3.6in
September 2007\\
\end{titlepage}
\vfill
\eject
\section{Introduction}
Branes in curved backgrounds are well-studied objects in string theory, M theory, and also in studies of four and five dimensional black holes. They provide insight into gauge-gravity correspondence, can be regarded as excitations about some general curved background and can be used to identify the non-perturbative states of strings in lower dimensions in different compactification schemes \cite{BESETO, BFSNTD, DUST, DUHOINST, BRPE, JN, FF, JMSS, PRT, Douglas, Tseytlin, CT, CVTS, CGJM, HT, T, JHS, SCHWARZ, Hora, BBS, W}. Such branes also figure in brane-world like scenarios where the universe is looked upon as a brane in a higher dimensional spacetime (\cite{IM}, \cite{PRT} and references therein). In particular, the study of time dependent backgrounds in the framework of string theory has attracted considerable attention for some time now, although the understanding reached so far is far from satisfactory. However, notwithstanding the very many challenges that beset the task, it is important that the subject be pursued vigorously for the obvious dividends it is likely to pay, such as - an understanding of time as an emergent phenomenon, a deeper insight into cosmological and black hole singularities, a realization of dS/CFT correspondence, a much-desired contact with experiment through observational cosmology, and the list goes on  \cite{TIME}. Thus, motivated by a desire to understand such interesting objects and backgrounds better, we study in this paper branes in various curved backgrounds. To that end, we develop a class of time dependent backgrounds and branes embedded therein, in eleven dimensional supergravity as well as in type-II string theories. This is done by a Wick rotation of the general configuration presented in \cite{ASS}. Wick rotation may at times lead to complex solutions; however, we are able to avoid such a situation due to the particular form of the reduced equations of motion obtained from eleven dimensional supergravity. The reduced equations of motion can be matched with dilaton-gravity equations in two dimensions derived from a certain two dimensional dilaton gravity action with a zero cosmological constant. Because of this rather specific correspondence to 2-d gravity, the equations happen to be completely integrable. We then solve one of the reduced equations of motion to obtain our time dependent solutions. The solutions so developed are then examined in their asymptotic and near-horizon limits and are also used to generate other solutions in type II theories. \\

The paper is organized as follows. In section 2, we review the exact, flat membrane, solution of the supergravity field equations and its extension to a class of curved membrane solutions. The reduced equations of motion are also presented for use in later sections. In section 3 we analytically continue the curved membrane solution of the previous section and, then, we go on to solve the reduced equations of motion to obtain time dependent branes in eleven dimensional supergravity. The brane solutions so obtained are then reduced to string solutions in type-IIA theory by a dimensional reduction along a worldvolume direction. Other brane solutions, such as a $D3$ brane, are then obtained by a succession of S and T-duality transformations in type-II theories. The asymptotic and the near-horizon geometries of all the solutions are also looked into and connection with $AdS_n\times S^{(d+1-n)}, (\,n=\,4,5;\: d=\,10,9)$ or $Milne\times R^p$ background is established. We end this section by giving, as an aside, a static, space dependent solution developed analogously to its time dependent counterpart. Finally, section 4 summarizes our results and mentions some possible extensions of our work.

\section{Review of a class of static, curved membrane solutions}
\noindent
As a preamble to our work, we quickly review the flat membrane solution of the supergravity field equations \cite{DUST} and its extension to a class of curved membrane solutions \cite{ASS}. Following \cite{DUST} we proceed by making an ansatz for the $D=11$ gauge fields $g_{MN}$ and $A_{MNP}$ ($ M, N = 0,1,\ldots,10 $) in conformity with the most general three-eight split invariant under $P_3\times SO(8)$ where $P_3$ and $SO(8)$ are the $D = 3$ Poincare group and the $D = 8$ rotation group respectively. The $D=11$ coordinates are also split as :
 $x^M=(x^\mu,y^m)$,
where $\mu=0,1,2$ and $m=3,\ldots,10$.

The metric is :
\begin{equation}
ds^2 = e^{2\tilde{A}}\,{\eta}_{{\mu}{\nu}}\,dx^{\mu}dx^{\nu} 
+e^{2\tilde{B}}\,{\delta}_{m n}\,dy^m dy^n,
\label{}
\end{equation}
and the three-form gauge field is :
\begin{equation}
A_{{\mu}{\nu}{\rho}} =\pm{1\over{{}^3g}}\,\epsilon_{{\mu}{\nu}{\rho}}\,e^{\tilde{c}},
\label{}
\end{equation}
where $^{3}g$ is the determinant of $g_{\mu\nu}$ and 
$\epsilon_{\mu\nu\rho}$ is the three dimensional Levi-civita tensor.
All other components of $A_{MNP}$ and all components of the
gravitino $\psi_M$ are set to zero. Invariance under $P_3\times SO(8)$ demands that the arbitrary functions $\tilde{A},\tilde{B}$ and $\tilde{C}$ depend on $y^m$ only through $r= \sqrt{(y^m)^2}$.

If now we require the field configurations (1) and (2) to preserve some unbroken supersymmetry, then there must exist Killing spinors $\epsilon$ satisfying 
\begin{equation}
{\tilde{D}_M}\epsilon = 0,
\label{}
\end{equation} 
where
\begin{equation}
{\tilde{D}_M}= \partial_M+\frac{1}{4}{\omega_M}^{AB}\,\Gamma_{AB}
-\frac{1}{288}\,({\Gamma^{PQRS}}_M +8\,\Gamma^{PQR}{{\delta}^S}_M)
F_{PQRS},
\label{}
\end{equation}

with $F_{MNPQ}=4\,{\partial_{[M}A_{NPQ}]}$.
Here ${\omega_M}^{AB}$ are the spin connections and $\Gamma_A$
are the $D=11$ Dirac matrices satisfying $\{\Gamma_A,\Gamma_B\} = 2\eta_{AB}$.
A,B refer to $D=11$ tangent space, $\eta_{AB}=diag(-,+,\ldots ,+)$
and $\Gamma_{AB\ldots C}= \Gamma_{[A}\Gamma_B \ldots \Gamma_{C]}$.
We then make a three-eight split:
$\Gamma_A= (\gamma_\alpha\otimes\Gamma_9, I \otimes{\Sigma}_a)$,
where $\gamma_\alpha$ and $\Sigma_a$ are the $D=3$ and $D=8$ Dirac 
matrices respectively and $\Gamma_9=\Sigma_3 \ldots \Sigma_{10}$.

The most general spinor field consistent with $P_3\times SO(8)$ assumes the form
\begin{displaymath}
\epsilon(x,y)=\epsilon \otimes \eta(r),
\end{displaymath}
where $\epsilon$ is a constant spinor of $SO(1,2)$ and $\eta$ is an $SO(8)$ spinor.
Further calculation shows that eq.\,(3) admits two non-trivial solutions $(1\pm\Gamma_9)\,\eta=0$, where the $\pm$ signs correspond to those in eq.\,(2),
\begin{displaymath}
\eta=e^{-{\frac{\tilde {C}}{6}}}\,\eta_0,
\end{displaymath}
where $\eta_0$ is a constant spinor, and
\begin{eqnarray*}
\tilde A &=& \frac{1}{3}\tilde C,\nonumber\\
\tilde B &=& -\frac{1}{6}\tilde C+ const.
\end{eqnarray*}
\noindent
In each solution of (3), one half of the maximal poissible rigid supersymmetry survives.
Thus we have seen that the requirement of some unbroken supersymmetry ( here, half ) reduces the arbitrary functions to one, namely, $\tilde{C}$. It now remains to determine $\tilde{C}$. To that end, we substitute the ansatz (eqs.\,1,\,2) into the field equations which follow from the action
\begin{equation}
S_G=\int d^{11} x \,{\cal L}_G,
\label{}
\end{equation} 
where ${\cal L}_G$ is the supergravity lagrangian whose bosonic sector is given by
\begin{eqnarray}
\kappa^2 \,{\cal L}_G &=&\frac{1}{2}{\sqrt{-g}}\,R-\frac{1}{96}\sqrt{-g}\,
F_{MNPQ}\,F^{MNPQ} \nonumber \\
&+&\frac{1}{2(12)^4}\,\epsilon^{MNOPQRSTUVW}\,F_{MNOP}\,F_{QRST}\,A_{UVW}.
\label{}
\end{eqnarray}
The three form field equation is given by
\begin{equation}
\partial_M\,(\sqrt{-g}\,F^{MUVW})+\frac{1}{1152}\,\epsilon^{UVWMNOPQRST}\,
F_{MNOP}\,F_{QRST} = 0.
\label{}
\end{equation}
The substitution yields $\delta^{mn}\,\partial_m \partial_n\,e^{-\tilde{C}} = 0$. On imposing the boundary condition that the metric be asymptotically Minkowskian, one gets $e^{-\tilde {C}}=1+\frac{K}{r^6}$, $r > 0$, where $K$ is a constant.
Thus we get,
\begin{eqnarray}
ds^2 &=& (1+\frac{K}{r^6})^{-\frac{2}{3}}\,\eta_{\mu\nu}\,dx^{\mu}dx^{\nu}
+(1+\frac{K}{r^6})^{\frac{1}{3}}\,\delta_{mn}\,dy^m dy^n,\nonumber\\
A_{\mu\nu\rho} &=& \pm\,\frac{1}{^{3}g}\,\epsilon_{\mu\nu\rho}\,(1+\frac{K}{r^6})^{-1}.
\label{}
\end{eqnarray}
The above expressions blow up at $r = 0$ ; hence they are solutions to the field equations everywhere except at that point. To remedy the situation, we modify the supergravity action by adding a source term at $r = 0$. The combined supergravity and membrane equations follow from the action  
\begin{equation}
S=S_G+S_M,
\label{} 
\end{equation}
where 
\begin{equation}
S_M = T\int d^3 \xi (-\frac{1}{2}\,\sqrt{-\gamma}\,\gamma^{ij}\partial_i
X^M \partial_j X^N g_{MN}+\frac{1}{2}\,\sqrt{-\gamma}\pm\frac{1}{3!}
\epsilon^{ijk}\,\partial_i X^M \partial_j X^N \partial_k X^P A_{MNP}),
\label{}
\end{equation}
where $T$ is the membrane tension. The Einstein equations are now
\begin{equation}
R_{MN}-\frac{1}{2}\,g_{MN}R=\kappa^2 T_{MN},
\label{}
\end{equation}
where,
\begin{eqnarray}
\kappa^2 T_{MN} = \frac{1}{12}\,(F_{MPQR}\,F_N^{PQR}-\frac{1}{8}\,g_{MN}
F_{PQRS}\,F^{PQRS}) \nonumber \\
-\kappa^2 T \int d^3 \xi \sqrt{-\gamma}\,\gamma^{ij}\,
\partial_i X_M\, \partial_j X_N \,\frac{\delta^{11}(x-X)}{\sqrt{-g}},
\label{}
\end{eqnarray}
while the three form equation is:
\begin{eqnarray}
\partial_M\,(\sqrt{-g}\,F^{MUVW})+\frac{1}{1152}\,\epsilon^{UVWMNOPQRST}
F_{MNOP}\,F_{QRST}= \nonumber \\
\pm 2\kappa^2 T \int d^3 \xi\, \epsilon^{ijk}\,\partial_i X^U 
\partial_j X^V \partial_k X^W \,\delta^{11}(x-X).
\label{}
\end{eqnarray}
In addition, we have the membrane field equations:
\begin{eqnarray}
& &\partial_i\,(\sqrt{-\gamma}\,\gamma^{ij}\,\partial_j X^N g_{MN})+
\nonumber \\
& &\frac{1}{2}\sqrt{-\gamma}\,\gamma^{ij}\,\partial_i X^N \partial_j X^P
\partial_M g_{NP}\,\pm \frac{1}{3!}\,\epsilon^{ijk}\,\partial_i X^N
\partial_j X^P \partial_k X^Q F_{MNPQ}=0,
\label{}
\end{eqnarray}
\begin{equation}
\gamma_{ij}=\partial_i X^M \partial_j X^N g_{MN}.
\label{}
\end{equation}
It is easy to verify that the correct source term is obtained by the
the static gauge choice $X^\mu =\xi^\mu$ and $Y^m$ =const, provided
$K=\frac{\kappa^2 T}{3\Omega_7}$ where $\Omega_7$ is the volume of
$S^7$.

Inspired by the results of \cite{DUST} , Bhattacharya et al give in \cite{ASS} a class of curved membrane  solutions using an ansatz which is a variation of the one in \cite{DUST}. The new ansatz is 
\begin{eqnarray}
ds^2 = e^{2A}[-(dx^0)^2+\sqrt{f}\,(dx^a)^2]+e^{2B}(dy^m)^2, 
\nonumber\\ \mbox{and}\ 
A_{\mu\nu\rho}=\pm\frac{1}{^{3}g}\,\epsilon_{\mu\nu\rho}\,e^C,
\nonumber\\ \mbox{with}\ 
e^{2A} =e^{2\tilde{ A}}F^e , e^{2B} =e^{2\tilde{B}}g^b,
e^C =e^{\tilde {C}}\chi^c , r^6 =h^d{\tilde{r}}^6,\nonumber\\ 
\tilde r=\sqrt{(y^m)^2}.
\label{}
\end{eqnarray}
$\tilde{A},\tilde{B},\tilde{C}$ are functions of $r$ and $F,g,\chi,f,h$ are 
functions of $x^a$. Here $\tilde{A},\tilde{B},\tilde{C}$ are related
as before viz.,$\tilde A=\frac{1}{3}\tilde C$, 
$\tilde B=-\frac{1}{6}\tilde C$+ const.
 
Here, the presence of the conformal factor $f$, which depends purely on the spatial directions $x^{1,2}$, introduces a non-vanishing curvature on the worldvolume of the membrane. In this context, coming to the question of preservation of supersymmetry, we notice that although for specific choices of $h$ and $f$ it is possible to retain some supersymmetry ( 1/2 or, 1/4 SUSY, for example ), for a generic choice of these factors supersymmetry is completely broken. To discuss these solutions in more detail we observe that, as before, the three-form field eq.\,(13) leads to 
\begin{equation}
e^{-\tilde {C}}=1+\frac{K}{r^6},
\label{}
\end{equation}
under the condition
\begin{equation}
F^{-\frac{3}{2}e}\,g^{3b}\,\chi^c f^{-\frac{1}{2}}\,h^{-d}=1.
\label{}
\end{equation} 
Similarly, the membrane field equation\,(14) yields the condition
\begin{equation}
F^{-{\frac{3}{2}}e}\,f^{-\frac{1}{2}}\,\chi^c  = 1.
\label{}
\end{equation}
The Einstein equation takes the form
\begin{equation}
R_{MN} = \kappa^2(T_{MN}-\frac{1}{9}\,g_{MN}T).
\label{}
\end{equation}
The ansatz is indeed a solution to the equations of motion provided we set
\begin{eqnarray}
F=g=h,\nonumber\\ \mbox{and,}\ 
2d= -3e = 6b. 
\label{set conds}
\end{eqnarray}
Also of interest are the following pair of equations obtained from eq.\,(21) :
\begin{eqnarray}
R_{ab}(\sqrt{f}F^e)+\frac{3e}{2}\nabla_a\nabla_b\ln F-\frac{3e^2}
{4}\nabla_a\ln F\nabla_b\ln F = 0,\nonumber\\
\nabla^2(e^{-{\frac{3e}{2}}\ln F}) = 0.
\label{}
\end{eqnarray}
Here $R_{ab}(\sqrt{f}F^e)$ denotes the Ricci tensor components for the conformal metric $g_{ab} = \sqrt{f}F^e\delta_{ab}$. 
Now, for two such metrics related by a conformal transformation, it follows from the standard rules [\,see appendix D,\cite{Wald}\,] 
 \begin{displaymath}
R_{ab} = {-{1\over4}}({f^{,\,c}\over{f}})\:_{,\,c}\;\tilde{g}_{ab} -{{e\over2}}\,({F^{,\,c}\over{F}})\:_{,\,c}\;\tilde{g}_{ab}, 
\end{displaymath}
and,
\begin{equation}
R = -{1\over{\sqrt{f}\;F^e\;}}[\;{1\over2}\;({f^{,\,c}\over{f}})\:_{,\,c}\; +\:e\;({F^{,\,c}\over{F}})\:_{,\,c}\;].
\label{wald}
\end{equation}
\noindent
Eq.\,(\ref{wald}) may be contracted to yield 
\begin{equation}
R\, -\, {3e\over2\sqrt{f}}\;{1\over{F^{e+1}}}\;[\; {({e\over2}+1)} \;{({F^{,\,c})\;(F_{,\,c})}\over{F}}\:-\:(F^{,\,c})\:{,\,c}\;] = 0.
\label{contracted wald}
\end{equation}
\noindent
Again, defining $\phi=\f{3e}{2}\ln F$, we can rewrite eq.\,(22) as:
\begin{eqnarray*}
R_{ab}(\sqrt{f}F^e)+\nabla_a\nabla_b\phi-\frac{1}{3}\,\nabla_a\phi\,\nabla_b\phi = 0,\nonumber\\
\nabla^2\,e^{-\phi}= 0.
\label{}
\end{eqnarray*}
The above equations match with the equations obtained from the
following two dimensional dilaton gravity action\cite{Cadoni}
provided $k=-\f{1}{2}$ and the cosmological term $\lambda$ 
is set to zero.\\
\begin{displaymath}
S = -\int_M\sqrt{g}\,e^{-2\phi}[R+\f{8k}{k-1}\,(\nabla\phi)^2+\lambda^2]\,-2\int_{\partial M}\,e^{-2\phi}K,
\label{}
\end{displaymath}
\noindent
where $K$ is the trace of the second fundamental form, $\partial M$  is the 
boundary of  $M$ and $k$ is a parameter taking values $|k|\leq 1$.\\
A necessary condition that the ansatz satisfies the supergravity equations of motion in eleven dimensions is :
\begin{eqnarray}
\frac{F_{,ab}}{F} + (d-1)\frac{F_{,a}F_{,b}}{F^2} - 
\frac{1}{4}\,\frac{1}{Ff}\,(f_{,a}F_{,b}+f_{,b}F_{,a}-f_{,c}F_{,c}\,\delta_{ab})=0,    \quad \quad  \quad  \quad \quad (a)
\nonumber\\
F_{,aa}+(4b+\frac{e}{2}-1)\,\frac{{F_{,a}}^2}{F}=0. \quad \quad \quad     \quad     \quad                (b)
\label{2d cond}
\end{eqnarray}
Finally, using (17) and (21) in (16), one gets the metric and the three form as :
\begin{eqnarray}
ds_{11}^2 &=& (F^d+\frac{K}{{\tilde{r}}^6})^{-\frac{2}{3}}\,
[-(dx^0)^2+\sqrt{f}\,(dx^a)^2]\nonumber\\
&+&(F^d+\frac{K}{{{\tilde{r}}^6}})^{\frac{1}{3}}\,(dy^{m})^2, \nonumber\\
A_{\mu \nu \rho} &=& \pm \frac{1}{^{3}g}\,\epsilon_{\mu \nu \rho}
(F^d+\frac{K}{{{\tilde{r}}^6}})^{-1}\,\sqrt{f}.
\label{}
\end{eqnarray}
We now go on to solve explicitly for $f(=g=h)$ and $F$ in order to construct various brane solutions.

\section{Time dependent solutions}
\noindent
To introduce time dependence into our solution, we perform a double analytic continuation with
\begin{eqnarray*}
                         x^0 \rightarrow i{x^2},\\
                         x^2 \rightarrow -{ix^o},
\end{eqnarray*}
reducing the metric and the three-form in (26) to the following forms :
\begin{eqnarray}
ds_{11}^2 &=& (F^d+\frac{K}{{\tilde{r}}^6})^{-\frac{2}{3}}\:[-{\sqrt{f}}\:{({dx^0})^2}+{\sqrt{f}}\:{({dx^1})^2}+{({dx^2})^2}]\nonumber \\ &+&(F^d+\frac{K}{{{\tilde{r}}^6}})^{\frac{1}{3}}\:{({dy^m})^2},\nonumber\\
A_{{\mu}{\nu}{\rho}} &=&\pm({1\over{{}^3g}}\,\epsilon_{{\mu}{\nu}{\rho}})\,(F^d+\frac{K}{{{\tilde{r}}^6}})^{-1}\:\sqrt{f}.
\label{analytic cont}
\end{eqnarray}
We note in this context that as the two coordinates $x^0$ and $x^2$ involved in the analytic continuation both belong to the worldvolume directions, the reality of the metric and the three-form field is still maintained. This would, however, not be the case if one of the coordinates was longitudinal and the other, transverse. Furthermore, $f\,(=g=h)$,\,$\chi, {\rm and}, F$, which were earlier functions of $x^1$ and $x^2$, now become functions of $x^0$ and $x^1$. And, in a purely time-dependent situation, we can take them all to be functions of $x^0$ only.

Next, we give an explicit solution of (\ref{2d cond}) with a view to obtaining a time-dependent brane solution. The general index structure of (\ref{2d cond}) allows us to reduce, in the time-dependent case, eq.\,(25) to :
\begin{equation}
\ddot{F} + {(d-1)} { \dot{F}^2\over{F}} =  0.
\label{time}
\end{equation}
Now, to solve this\footnote{originating from a suggestion by S.Mukherji}, we propose :
\begin{equation}
F = e^z.
\label{ansatz}
\end{equation}
This, when substituted in (\ref{time}), yields,
\begin{equation}
\ddot{z} + d\, \dot{z}^2 =  0.
\label{}
\end{equation}
\noindent
The above equation is then solved for z and the solution is substituted back in (\ref{ansatz}) to obtain the following solution for F :
\begin{equation}
F = F_0 \;(\,d \:t -c)^{1\over\tilde{d}},
\label{soln F}
\end{equation}
\noindent
where $F_0$ and c are integration constants.

It now remains to determine the unknown function $f$ in (\ref{analytic cont}). In a purely time dependent situation, (\ref{contracted wald}) is further simplified, by using (\ref{time}) and (\ref{set conds}), to, 
\begin{equation}
R - {2\over3}\: {\,d^2}\: {1\over{\sqrt{f}\;F^e\;}}\: {F_{,t}^2\over {F^2}} = 0.
\label{above}
\end{equation}
\noindent
Using (\ref{wald}),\,(\ref{set conds}), and\,(\ref{time}) in (\ref{above}), one can, then, readily show that the solution for $f$ is : 
\begin{equation}
f = B\,e^{at},
\label{soln f}
\end{equation}
\noindent
where a and B are integration constants.

Finally, using (\ref{soln F}) and (\ref{soln f}) we can now recast (\ref{analytic cont}) into the form :
\begin{eqnarray}
ds_{11}^2 &=& (F_0^{d}\:(d\:t-c)+
{K\over{\tilde{r}^6}})
^{-{2\over3}}\:
[-B^{1\over2}\,e^{at\over2}\:(dt)^2+B^{1\over2}\,e^{at\over2}\:(dx^1)^2+(dx^2)^2]\nonumber\\ 
&+& (F_0^{d}\:(d\:t-c)+
{K\over{\tilde{r}^6}})
^{1\over3}\:(dy^m)^2,\nonumber\\
A_{{\mu}{\nu}{\rho}} &=&\pm({1\over{}^3g}\,\epsilon_{{\mu}{\nu}{\rho}})\,
(F_0^{d}\:(d\:t-c)+{K\over{\tilde{r}^6}})^{-1}\:
B^{1\over2}\,e^{at\over2}.
\label{M-brane soln}
\end{eqnarray}
\noindent
The relationship with the flat-brane solution is thus implied through a replacement of the constant part of the Green's function in (17) by a time dependent one. Eq.\,(34) can, however, be also expressed in the following simpler form :
\begin{eqnarray}
ds_{11}^2 &=& (\tilde{t}+
{\tilde{K}\over
{\tilde{r}^6}})
^{-{2\over3}}\:
[- K_0\,e^{{\tilde{a}\tilde{t}}}\:(d{\tilde{t}})^2+e^{{\tilde{a}\tilde{t}}}\:(d{\tilde{x}}^1)^2+(d{\tilde{x}}^2)^2]\nonumber\\ 
&+&(\tilde{t}+
{\tilde{K}\over
{\tilde{r}^6}})
^{1\over3}\:(d{\tilde{y}}^m)^2,\nonumber\\
\tilde{A}_{012} &=&\pm\, M\,
 (\tilde{t}+
{\tilde{K}\over
{\tilde{r}^6}})
^{-1}\:e^{{\tilde{a}\tilde{t}}},
\label{re-scaled M-brane soln}
\end{eqnarray}
\noindent
where we have made use of the following substitutions in order to get rid of the redundant parameters contained in (\ref{M-brane soln}) :
\begin{eqnarray}
\tilde{K} &=&\frac{K}{F_0^{d}}\:;\;\tilde{a} = \frac{a}{2d}\:;\;K_0 =F_0^{-\frac{2d}{3}}\,B^{\frac{1}{2}}\,e^{\frac{ac}{2d}}\,d^{-2}\:;\;M =F_0^{-\frac{d}{3}}\,B^{\frac{1}{4}}\,e^{\frac{ac}{4d}}\,d^{-1}\:;\;\nonumber\\
\tilde{t} &=& d\,t-c\:;\;
\tilde{x}^1 = F_0^{-\frac{d}{3}}\,B^{\frac{1}{4}}\,e^{\frac{ac}{4d}}\,x^1\:;\;
\tilde{x}^2 = F_0^{-\frac{d}{3}}\,x^2\:;\;
\tilde{y}^m = F_0^{\frac{d}{6}}\,y^m.
\label{re-scaled variables}
\end{eqnarray}
\noindent
Eq.\,(\ref{M-brane soln}), or its counterpart in the re-scaled variables, eq.\,(\ref{re-scaled M-brane soln}), constitutes one of our main results as it gives us an exact time dependent M2 brane solution of 11-d supergravity equations of motion. The two forms of the solution, nonetheless, enjoy different properties (for the choice of $c=-1$, $d=0$), and so we retain them both. Now, to understand the properties of such a brane solution, one needs to analyze geometric quantities such as curvature and other field strengths. However, owing to the complexity of the above solution, we study it analytically in its asymptotic and near-horizon limits only.\\

(a) Asymptotic limit : ${\tilde{r}\,\rightarrow \infty} $: In this limit the soln.\,(\ref{M-brane soln}) goes to :
\begin{eqnarray}
ds_{11}^2 &=&  (F_0^{d}\:(d\:t-c))^{-{2\over3}}\:
[-B^{1\over2}\,e^{at\over2}\:(dt)^2+B^{1\over2}\,e^{at\over2}\:(dx^1)^2+(dx^2)^2]\nonumber\\
&+&(F_0^d\:(d\:t-c))^{1\over3}\:(dy^m)^2,\nonumber\\
A_{{\mu}{\nu}{\rho}} &=&\pm({1\over{}^3g}\,\epsilon_{{\mu}{\nu}{\rho}})\,
(F_0^{d}\:(d\:t-c))^{-1}\:B^{1\over2}\,e^{at\over2}.
\label{11-d asym lt}
\end{eqnarray}
\noindent
It follows from (35) that the same limit, in terms of the re-scaled varibles introduced in (\ref{re-scaled variables}), reads as :
\begin{eqnarray}
ds_{11}^2 &=& \tilde{t}^{-{2\over3}}\:
[-\,K_0\,e^{{\tilde{a}\tilde{t}}}\:(d\tilde{t})^2+e^{{\tilde{a}\tilde{t}}}\:(d{\tilde{x}}^1)^2+(d{\tilde{x}}^2)^2]+\tilde{t}^{1\over3}\:(d{\tilde{y}}^m)^2,
\nonumber\\
\tilde{A}_{012} &=&\pm\,M\,
\tilde{t}^{-1}\:
e^{{\tilde{a}\tilde{t}}},
\label{re-scaled 11-d asym lt}
\end{eqnarray}
\noindent
where M is as defined in (\ref{re-scaled variables}).\\
The Ricci scalar curvature for the above background is zero ($R=0$), and \,$R^2(\equiv{R^{\mu\nu}R_{\mu\nu}})$\,is:
\begin{equation}
R^2=N\,e^{-2{\tilde{a}\tilde{t}}}\:\tilde{t}^{-{2\over3}},\:
\label{R-squared}
\end{equation}
\noindent
where $N={\tilde{a}^2\over2\,K_0^2}$. This shows that the background has a space-like singularity at $\tilde{t} = 0$. The singularity, however, merits closer attention and, to that end, we proceed as follows.

The geodesic equation for the metric described by (\ref{re-scaled 11-d asym lt}) gives
\begin{eqnarray}
\frac{d^2\tilde t} {d\lambda^2} +\,\Gamma_{\tilde t \,\tilde t}^{\tilde t}\, (\frac{d \tilde t}{d \lambda})^2 +\,\sum_{i=1}^{10}\:\Gamma_{{\tilde x}^i \,{\tilde x}^i}^{\tilde t}\, (\frac{d {\tilde x}^i}{d \lambda})^2 &=& 0\,, \hspace{1cm} \Gamma_{\tilde t \,\tilde t}^{\tilde t}=(\frac{\tilde a}{2} - \frac{1}{3 \,\tilde t})
\quad \quad  (a) 
\nonumber\\
\frac{d^2{\tilde x}^i} {d\lambda^2} +\Gamma_{\tilde t \,{\tilde x}^i}^{{\tilde x}^i}\, 
(\frac{d \tilde t}{d \lambda})\,(\frac{d {{\tilde x}^i}}{d \lambda}) &=& 0\,,   
\quad \quad \quad \quad \quad \quad \quad \quad \quad \quad \quad  (b)
\label{}
\end{eqnarray}
\noindent
where $\lambda$ is the affine parameter for the geodesic. Since $(\frac{d {{\tilde x}^i}}{d \lambda})\,=0$ is a solution for eq.\,(40\,b), we make use of it in eq.\,(40\,a) to obtain
\begin{equation}
\frac{d^2\tilde t} {d\lambda^2} +\,\Gamma_{\tilde t \,\tilde t}^{\tilde t}\, (\frac{d \tilde t}{d \lambda})^2 = 0.
\label{geodesic eqn}
\end{equation}
\noindent
To solve (\ref{geodesic eqn}) we introduce the auxiliary functions $y(\tilde{t})$ and $p^{\prime}(\tilde{t})$ such that, 
\begin{eqnarray}
y(\tilde{t}) &\equiv& {d\tilde{t}\over d\lambda},\nonumber\\
{\rm and},\;
p^{\prime}(\tilde{t}) &\equiv&{\tilde{a}\over2} - {1\over{3\,\tilde{t}}}. 
\label{auxiliary functions}
\end{eqnarray}
\noindent
Eq.\,(\ref{geodesic eqn}), when expressed in terms of the above functions, becomes
\begin{equation}
{dy\over d\lambda}\,+\,p^{\prime}(\tilde{t})\,y^2 = 0.
\label{modified geodesic eqn}
\end{equation}
\noindent
Next, we solve (\ref{modified geodesic eqn}) to obtain the following expression for $\lambda$ in terms of $p(\tilde{t})$ :
\begin{equation}
\lambda = {\rm const}\,\int e^{p(\tilde{t})}\,d\tilde{t}.
\label{lambda soln 1}
\end{equation}
\noindent
En route to (\ref{lambda soln 1}) we also obtain the following intermediate step which we record here for use in the calculation of gauge invariant curvature below :
\begin{equation}
y(\tilde{t}) \equiv {d\tilde{t}\over d\lambda} = {\rm const}\:e^{-p}.
\label{intermediate eqn}
\end{equation}
\noindent
Again, it follows from (\ref{auxiliary functions}) that the solution for $p(\tilde{t})$ is
\begin{equation}
p(\tilde{t}) = {\tilde{a}\,t\over2}+ln\,\tilde{t}^{-{1\over3}}\,+\,const.
\label{p soln}
\end{equation}
\noindent
With the explicit expression for $p(\tilde{t})$ obtained above, (\ref{lambda soln 1}) can now be integrated, by expanding $e^{\tilde{a}\,t\over2}$ in a series, to obtain : 
\begin{equation}
\lambda = {\rm const} \sum_{n=0}^\infty \,{1\over n!}\,({\tilde{a}\over 2})^n\,{3\over{3\,n+\,2}}\:\tilde{t}^{{3\,n\,+\,2}\over3}\;+\,{\rm const}.
\label{affine parameter}
\end{equation}
\noindent
The above series converges for $\tilde{t}\geq0$ and the appearance of the two integration constants on the r.h.s. of (\ref{affine parameter}) is a consequence of the second order nature of (\ref{geodesic eqn}).\\
Now, the gauge invariant curvature defined along the geodesic is 
\begin{eqnarray}
R_{\lambda\lambda}(\lambda) \equiv R_{\mu\nu}\,{dx^{\mu}\over d{\lambda}}\,{dx^{\nu}\over d{\lambda}} &=& R_{00}\,({d\,\tilde{t}\over d{\lambda}})^2,  \quad \quad {\rm where}\: R_{00}=-{{\tilde{a}\over2\,\tilde{t}}},\nonumber\\
&=&{\rm const} (-\frac{ \tilde a}{ 2 \, \tilde t})\, e^{- 2 \,p(\tilde t)}, \nonumber\\
&=&{\rm const}\frac{e^{ - \tilde  a \,\tilde t}}{\tilde t^{\frac{1}{3}}},
\label{gauge invariant curvature}
\end{eqnarray}
\noindent
where we have made use of\,(\ref{intermediate eqn},\,\ref{p soln}). 
This blows up at $\tilde{t} = 0$, as does $R^2$ (\ref{R-squared}), and (\ref{affine parameter}) shows that the corresponding value of $\lambda$ is finite.  The spacetime, therefore, is singular as it has a curvature singularity at finite affine parameter, making it geodesically incomplete. The background (\ref{re-scaled 11-d asym lt}) is, thus, cosmological and the full solution (\ref{M-brane soln}/\ref{re-scaled M-brane soln}) can be interpreted as an M2 brane in such a cosmological background. Furthermore, our background (\ref{re-scaled 11-d asym lt}) can also be re-written in terms of scale factors $a_i(t)$'s  as in\cite{Veneziano} and it will be interesting to see what kind of universe this leads to. We, however, at this stage, leave the exercise of finding the $a_i(t)$'s explicitly.

On the other hand, for $d=0$, $c=-1$, the asymptotic limit of the full solution (eq.\,\ref{11-d asym lt}) becomes : 
\begin{eqnarray}
ds_{11}^2 &=& [-B^{1\over2}\,e^{at\over2}\:(dt)^2+B^{1\over2}\,e^{at\over2}\:(dx^1)^2+(dx^2)^2]+(dy^m)^2,\nonumber\\
A_{{\mu}{\nu}{\rho}} &=&\pm({1\over{}^3g}\,\epsilon_{{\mu}{\nu}{\rho}})\,
\:B^{1\over2}\,e^{at\over2}.
\end{eqnarray}
\noindent
Now, with a coordinate transformation of the form
\begin{displaymath}
\tilde{t} = {4\over{a}}\: B^{1\over4}\,e^{a{t}\over4},
\end{displaymath}
eq.\,(49) reduces to :
\begin{eqnarray}
ds_{11}^2 &=&[-(d{\tilde{t}})^2+{({a\over4})^2(\tilde{t})^2(dx^1)^2}]+ (dx^2)^2
+(dy^m)^2,\nonumber\\
\tilde{A}_{012}  &=&\pm\:({a\over4})\,{\tilde{t}}.
\end{eqnarray}
The spacetime described by (50) can be looked upon as a direct product of a $d-1$ dimensional flat Euclidean space $R^{d-1}$ with a two dimensional spacetime $M_c$\, (\,i.e \,$R^{d-1}\times\,M_c$\,), with the line element 
\begin{displaymath}
ds_{2}^2 =-(d{\tilde{t}})^2+{({a\over4})^2(\tilde{t})^2(dx^1)^2}.
\end{displaymath}
Alternatively, if we re-express the metric in terms of Minkowski light-cone coordinates 
\begin{displaymath}
x^{\pm} ={{1\over{\sqrt{2}}}\,{\tilde{t}}\,e^{{\pm}\,{a\over4}\,{x^1}}},
\end{displaymath}
such that 
\begin{eqnarray}
ds_{11}^2 &=&-2\,dx^+\,dx^-\,+ (dx^2)^2+(dy^m)^2,\nonumber\\
A^{\prime}_{+-2} &=&\mp1,
\end{eqnarray}\\
then the above expression shows that  $M_c$ is flat with a constant 3-form.

Now, if we allow the identification $x^1\sim  x^1+2\pi$, $M_c$ essentially becomes what is known as the Milne spacetime. The Milne spacetime consists of a circle with a time-dependent radius. As t runs from $-\infty$ to $\infty$, the Milne universe starts out with an infinite radius, shrinks away, and, then, grows once more. The instant $t=0$, thus, corresponds to a big crunch / big bang like singularity. Away from $t=0$, the metric is locally flat and it has a curvature singularity at $t=0$, unless $a\over4$ equals a positive integer. The identification  $x^1\sim  x^1+2\pi$ implies the following identification for the light-cone coordinate :
\begin{displaymath}
x^{\pm} \sim e^{\pm\,2\pi\gamma}\,x^{\pm}.
\end{displaymath}
We can, therefore, define Milne space as a two-dimensional Minkowski space with identification through Lorentz boost. Strings in such backgrounds have been studied in \cite{NEK, BCKR, PB, BPR, BDPR, DP, Hiki}.

(b) Near-horizon limit : $ {\tilde{r}\,\rightarrow 0}$ : In this case the soln. (eq.\,\ref{re-scaled M-brane soln})  yields
\begin{eqnarray}
ds_{11}^2 &=& 
\tilde{K}^{-{2\over3}}\:\tilde{r}^4\:
[-K_0\,e^{\tilde{a}\tilde{t}}\:(d{\tilde{t}})^2+e^{\tilde{a}\tilde{t}}\:(d{\tilde{x}}^1)^2+(d{\tilde{x}}^2)^2]+\tilde{K}^{1\over3}\:\tilde{r}^{-2}\:(d{\tilde{y}}^m)^2,\nonumber\\
A_{{\mu}{\nu}{\rho}} &=&\pm({M\over{}^3g}\,\epsilon_{{\mu}{\nu}{\rho}})\,\tilde{K}^{-1}\;\tilde{r}^6\:
e^{{\tilde{a}\tilde{t}}}.
\label{}
\end{eqnarray}
\noindent
For $\tilde{a}=0$, the near-horizon geometry of the M2-brane is seen to be pure $AdS_4\times S_7$. And, for $\tilde{a}\neq 0$, the AdS boundary
\begin{displaymath}
ds_{boundary}^2 = [-K_0\,e^{\tilde{a}\tilde{t}}\:(d{\tilde{t}})^2+e^{\tilde{a}\tilde{t}}\:(d{\tilde{x}}^1)^2+(d{\tilde{x}}^2)^2]
\label{}
\end{displaymath}
\noindent
can be attributed a $Milne\times R$ structure by introducing transformations and identifications in the spirit of the preceding discussion. The full solution (35), therefore, interpolates smoothly between an $AdS_4\times S_7$ space with a cosmological boundary and a time dependent, cosmological background at infinity (38).\\ 

We now perform a simultaneous dimensional reduction of spacetime and worldvolume to reduce our brane solution in eleven-dimensional supergravity to a string solution in type-IIA theory by identifying the eleventh spacetime dimension with the third dimension i.e. the $x^2$ direction of the worldvolume and, then, by compactifying it on a circle while simultaneously discarding the massive modes. We do this by following the dictionary for dimensional reduction given in \cite{BHO}.\\

Let us make the following ten-one split and denote all $D = 11 $ variables by a carat.
${{\hat{x}}^{\hat{a}}} = (\:x^M,\:x^2), \; M = 0,\,1,\,3,\,.......,10, \;\hat{a} = 0,\,1,\,2,\,3,\,.......,10.$  
The ten-dimensional fields, when expressed in terms of the eleven-dimensional ones, are as follows :
\begin{eqnarray}
g_{MN} &= &(\:{\hat{g}}_{{\hat{2}}{\hat{2}}}\:)^{1\over2}\,
(\:{\hat{g}}_{{\hat{M}}{\hat{N}}}\:-\:{{\hat{g}}_{{\hat{M}}{\hat{2}}}\:{\hat{g}}_{{\hat{N}}{\hat{2}}}}/\:{{\hat{g}}_{{\hat{2}}{\hat{2}}}})\nonumber\\
B_{MN}^{(1)} &=& {3\over2}\:{\hat{A}}_{{\hat{M}}{\hat{N}}{\hat{2}}}\:,\nonumber\\
A_{MNP} &=& {\hat{A}}_{{\hat{M}}{\hat{N}}{\hat{P}}}\:,\nonumber\\
A_M^{(1)}  & =& \frac{\hat g_{\hat M \,\hat 2}}{\hat g_{\hat 2 \, \hat 2}},\nonumber\\
\phi &=& {3\over4}\:log\,(\:{\hat{g}}_{{\hat{2}}{\hat{2}}}\:)\:.
\label{}
\end{eqnarray}
\noindent
Following the above prescription, we write down the $ D = 10 $ solution from the corresponding expression for $ D = 11 $ (eq.\,\ref{re-scaled M-brane soln}) : 
\begin{eqnarray}
ds_{10}^2 &=& (\tilde{t}+
{\tilde{K}\over
{\tilde{r}^6}})
^{-1}\:
[-\,K_0\,e^{{\tilde{a}\tilde{t}}}\:(d{\tilde{t}})^2+e^{{\tilde{a}\tilde{t}}}\:(d{\tilde{x}}^1)^2]+ (d{\tilde{y}}^m)^2,\nonumber\\
B_{01}^{(1)}  &=&\pm\,{3\over2}\,M\,(\tilde{t}+{\tilde{K}\over{\tilde{r}^6}})^{-1}\:
e^{{\tilde{a}\tilde{t}}},\nonumber\\
\phi &=& -{1\over2}\:log[\,(\tilde{t}+{\tilde{K}\over{\tilde{r}^6}})\,].
\label{10-d string soln}
\end{eqnarray}
\noindent
The asymptotic limit of the above solution is :
\begin{eqnarray}
ds_{10}^2 &=& \tilde{t}^{-1}\:[-\,K_0\,e^{{\tilde{a}\tilde{t}}}\:(d{\tilde{t}})^2+e^{{\tilde{a}\tilde{t}}}\:(d{\tilde{x}}^1)^2]+ (d{\tilde{y}}^m)^2,\nonumber\\
B_{01}^{(1)}  &=&\pm\,{3\over2}\,M\,\tilde{t}^{-1}\:e^{{\tilde{a}\tilde{t}}},\nonumber\\
\phi &=& -{1\over2}\:log[\,\tilde{t}\,].
\label{10-d asym string soln}
\end{eqnarray}
\noindent
\\
The solution (\ref{10-d string soln}) can be interpreted as  an F-string in a time dependent background (\ref{10-d asym string soln}).

Now, by an S-duality transformation of type-IIB theory\cite{PAPER} :
\begin{eqnarray}
                         \phi \:\rightarrow \:\phi^{\prime} = - \:\phi,\nonumber\\
                         B_{{\mu}{\nu}}^{(1)}\:\rightarrow\: B_{{\mu}{\nu}}^{(2)},\nonumber\\                        
                         B_{{\mu}{\nu}}^{(2)}\:\rightarrow\:- B_{{\mu}{\nu}}^{(1)},
\label{S duality}	                         
\end{eqnarray}
\noindent
where the superscripts `(1)' and `(2)' in (\ref{S duality}) refer to the NS-NS and the R-R fields respectively, one obtains the D-string solution in a time dependent background :
\begin{eqnarray}
ds_{10}^2 &=&(\tilde{t}+
{\tilde{K}\over
{\tilde{r}^6}})
^{-{1\over2}}\:
[-K_0\,e^{{\tilde{a}\tilde{t}}}\:(d{\tilde{t}})^2+e^{{\tilde{a}\tilde{t}}}\:(d{\tilde{x}}^1)^2]+\,(\tilde{t}+{\tilde{K}\over{\tilde{r}^6}})^{1\over2}\:
 (d{\tilde{y}}^m)^2,\nonumber\\
B_{01}^{(1)} &=& 0,\nonumber\\
B_{01}^{(2)}  &=&\pm\,{3\over2}\,M\,(\tilde{t}+{\tilde{K}\over{\tilde{r}^6}})^{-1}\:
e^{{\tilde{a}\tilde{t}}},\nonumber\\
\phi &=& {1\over2}\:log[\,(\tilde{t}+{\tilde{K}\over{\tilde{r}^6}})\,].
\label{}
\end{eqnarray}
\noindent
 
This may now serve as a starting point to generate other time dependent D-brane solutions; for example, one can obtain a D3 brane solution by applying two successive T-duality transformations along, say, $x^3$ and $ x^4$ directions respectively \cite{Breckenridge}, to obtain:
\begin{eqnarray}
ds_{10}^2 &=&(\tilde{t}+
{\tilde{K}\over
{\tilde{r}^4}})^{-{1\over2}}\:
[-K_0\,e^{{\tilde{a}\tilde{t}}}\:(d{\tilde{t}})^2+e^{{\tilde{a}\tilde{t}}}\:(d{\tilde{x}}^1)^2+\:(d{\tilde{x}}^3)^2+\:(d{\tilde{x}}^4)^2]\nonumber\\
&+&\,(\tilde{t}+{\tilde{K}\over{\tilde{r}^4}})^{1\over2}\: (d{\tilde{y}}^m)^2,\nonumber\\
A_{0134}^{+\,(4)}  &=&\pm\,{3\over2}\,M\,
 (\tilde{t}+
{\tilde{K}\over
{\tilde{r}^4}})
^{-1}\:e^{{\tilde{a}\tilde{t}}},\nonumber\\
\phi &=& 0. 
\label{}
\end{eqnarray}
\noindent
The asymptotic and the near-horizon limits now are :

(a) Asymptotic limit : ${\tilde{r}\,\rightarrow \infty} $ :
\begin{eqnarray}
ds_{10}^2 &=&\tilde{t}^{-{1\over2}}\:
[-K_0\,e^{{\tilde{a}\tilde{t}}}\:(d{\tilde{t}})^2+e^{{\tilde{a}\tilde{t}}}\:(d{\tilde{x}}^1)^2+\:(d{\tilde{x}}^3)^2+\:(d{\tilde{x}}^4)^2  ]+\,\tilde{t}^{1\over2}\: (d{\tilde{y}}^m)^2,\nonumber\\
A_{01{3}{4}}^{+\,(4)}   &=&\pm\,{3\over2}\,M\,\tilde{t}^{-1}\:
e^{{\tilde{a}\tilde{t}}},\nonumber\\
\phi &=& 0. 
\label{}
\end{eqnarray}
\noindent
As before, to explore the nature of curvature of this background, we compute the following:
\begin{eqnarray}
R&=&0,\nonumber\\
R^2(\equiv{R^{\mu\nu}R_{\mu\nu}})&=& N\,e^{-2{\tilde{a}\tilde{t}}}\:\tilde{t}^{-1}.
\label{}
\end{eqnarray}
\noindent
Thus, like its counterpart in the 11-d theory, this background too is cosmological because it has a space-like singularity at $\tilde{t}=0$ and can be shown to be geodesically incomplete.  

(b) Near-horizon limit : ${\tilde{r}\, \rightarrow 0}$ :
\begin{eqnarray}
ds_{10}^2 &=&\tilde{K}^{-{1\over2}}\:\tilde{r}^2\:
[-K_0\,e^{{\tilde{a}\tilde{t}}}\:(d{\tilde{t}})^2+e^{{\tilde{a}\tilde{t}}}\:(d{\tilde{x}}^1)^2+\:(d{\tilde{x}}^3)^2+\:(d{\tilde{x}}^4)^2  ]+\,\tilde{K}^{1\over2}\:\tilde{r}^{-2}\: (d{\tilde{y}}^m)^2,\nonumber\\
A_{0134}^{+\,(4)}   &=&\pm{3\over2}\,M\,\tilde{K}^{-1}\;\tilde{r}^4\:
e^{{\tilde{a}\tilde{t}}},\nonumber\\
\phi &=& 0.
\label{}
\end{eqnarray}
\noindent
We now make the following observations :\\
1.  Eq.\,(61) corresponds to an $AdS_5\times S^5$ with a cosmological boundary. 
The boundary metric for the AdS space
\begin{displaymath}
ds_{boundary}^2 = [-K_0\,e^{{\tilde{a}\tilde{t}}}\:(d{\tilde{t}})^2+e^{{\tilde{a}\tilde{t}}}\:(d{\tilde{x}}^1)^2+\:(d{\tilde{x}}^3)^2+\:(d{\tilde{x}}^4)^2] 
\label{}
\end{displaymath}
\noindent
is identified, by transformations and identifications very similar to the ones following eq.(49), with $Milne\times R^2$.\\ 
2.   For $\tilde{a} = 0$, the near-horizon limit (61) of the full metric (58) is pure $AdS_5\times S^5$ whereas the asymptotic counterpart (59) of the same (58) is time dependent. We, therefore, see that   the near-horizon geometry is insensitive to the asymptotic structure.\\ 
3.   The full solution (58) interpolates between an $AdS_5\times S^5$ at the horizon and a cosmological background (59) at infinity. 

Finally, as an aside, we solve, in a similar fashion, the reduced eqs.of motion (\ref{2d cond}) where   $f(=g=h), \chi,\, {\rm and}\, F$ are all functions of a worldvolume spatial coordinate, $z\,(\,\equiv{x^2})$, instead of time t. Then, by substituting the solutions for $f\,{\rm and}\,F$ into (26), we can write down a space-depedent solution
\begin{eqnarray}
ds_{11}^2 &=& (F_0^{d}\:(d\:z-c)+
{K\over{\tilde{r}^6}})
^{-{2\over3}}\:
[-(dt)^2+B^{1\over2}\,e^{az\over2}\:(dx^1)^2+B^{1\over2}\,e^{az\over2}\:(dz)^2]\nonumber\\ 
&+& (F_0^{d}\:(d\:z-c)+
{K\over{\tilde{r}^6}})
^{1\over3}\:(dy^m)^2,\nonumber\\
A_{{\mu}{\nu}{\rho}} &=&\pm({1\over{}^3g}\,\epsilon_{{\mu}{\nu}{\rho}})\,
(F_0^{d}\:(d\:z-c)+{K\over{\tilde{r}^6}})^{-1}\:
B^{1\over2}\,e^{az\over2},
\label{space dependent M-brane soln}
\end{eqnarray}
\noindent
the asymptotic limit of which is
\begin{eqnarray}
ds_{11}^2 &=& (F_0^{d}\:(d\:z-c))^{-{2\over3}}\: 
[-(dt)^2+B^{1\over2}\,e^{az\over2}\:(dx^1)^2+B^{1\over2}\,e^{az\over2}\:(dz)^2]\nonumber\\
&+&(F_0^{d}\:(d\:z-c))^{1\over3}\:(dy^m)^2,\nonumber\\
A_{{\mu}{\nu}{\rho}} &=&\pm({1\over{}^3g}\,\epsilon_{{\mu}{\nu}{\rho}})\,
(F_0^{d}\:(d\:z-c))^{-1}\:B^{1\over2}\,e^{az\over2}.
\end{eqnarray}
\label{}
\noindent
As in the time-dependent case, this background too may be shown to be Ricci flat with a time-like singularity, manifested through a singular $R^2(\equiv{R^{\mu\nu}R_{\mu\nu}})$ at $z=0$. Thus, we have a class of static, singular space-dependent solutions.

\section{Summary and discussion}
\noindent
To summarize, we have constructed time-dependent solutions in eleven dimensional supergravity and in type-II string theories, followed by an analysis of these solutions in their asymptotic and near-horizon limits. Although in this paper we have not attempted to understand the full solutions, the complexity of their structure, however, is likely to make the effort a rewarding one. The near-horizon limits of our time dependent solutions exhibit an $AdS_n\times S^{(d+1-n)}, (\,n=\,4,5;\: d=\,10,9)$ structure with a $Milne\times R^p, (\,p=\,1,2)$ boundary and their asymptotic counterparts provide us with examples of singular, cosmological backgrounds. The full solutions, therefore, can be looked upon as time dependent branes in such cosmological backgrounds. As a possible extension of our work, we suggest performing, in the near-horizon limit, a complimentary gauge theory study on the boundary to gain more insight into the AdS bulk. One can, furthermore, study the propagation of strings and branes as probes in such singular backgrounds with the hope of ``resolving" the singularities, or, at least, shedding more light on them. An understanding of the general structure of intersecting time dependent brane configurations may also prove to be worthwhile. Finally, one can similarly construct null solutions and study them in reference to pp waves and defect CFT's. 
\section{Acknowledgement}
\noindent
It is a pleasure to thank A. Kumar for suggesting the problem, based on his discussions with S. Mukherji, and guiding the work to completion. I would also like to thank S. Bhattacharyya, S. Mukhopadhyay, S. Sarkar, and, in particular, T. Dey for help at various stages of the work.

\end{document}